\documentclass[fleqn,usenatbib]{mnras}

\usepackage[T1]{fontenc}
\usepackage{ae,aecompl}
\usepackage[export]{adjustbox}
\usepackage{graphicx}
\usepackage{times}

\usepackage[pscoord]{eso-pic}
\usepackage{amsmath}
\usepackage{amssymb}

\newcommand{\msun}{{\,\rm M_\odot}}

\def\gsim{ \lower .75ex \hbox{$\sim$} \llap{\raise .27ex \hbox{$>$}} }
\def\lsim{ \lower .75ex \hbox{$\sim$} \llap{\raise .27ex \hbox{$<$}} }
\defcitealias{Viel05}{V05}
\defcitealias{Lovell14}{L14}

\title[Alternative DM mass functions]{The halo mass function in alternative dark matter models}

\author[M.~R.~Lovell]{
M.~R.~Lovell$^{1}$\thanks{E-mail:lovell@hi.is}
\\
$^{1}$Center for Astrophysics and Cosmology, Science Institute, University of Iceland, Dunhagi 5, 107 Reykjavik, Iceland\\
}

\date{Accepted 2020 January 2. Received 2019 December 16; in original form 2019 November 29}

\pubyear{2020}

\begin{document}
\label{firstpage}
\pagerange{\pageref{firstpage}--\pageref{lastpage}}
\maketitle

\begin{abstract}

\noindent The claimed detection of large amounts of substructure in lensing flux anomalies, and in Milky Way stellar stream gaps statistics, has lead to a step change in constraints on simple warm dark matter models. In this study we compute predictions for the halo mass function both for these simple models and also for comprehensive particle physics models of sterile neutrinos and dark acoustic oscillations. We show that the mass function fit of Lovell~et~al. underestimates the number of haloes less massive than the half-mode mass, $M_\rmn{hm}$ by a factor of 2, relative to the extended Press-Schechter (EPS) method. The alternative approach of applying EPS to the Viel~et~al. matter power spectrum fit instead suggests good agreement at $M_\rmn{hm}$ relative to the comprehensive model matter power spectra results, although the number of haloes with mass $<M_\rmn{hm}$ is still suppressed due to the absence of small scale power in the fitting function. Overall, we find that the number of dark matter haloes with masses $<10^{8}\msun$ predicted by competitive particle physics models is underestimated by a factor of $\sim2$ when applying popular fitting functions, although careful studies that follow the stripping and destruction of subhaloes will be required in order to draw robust conclusions.   

\end{abstract}

\begin{keywords}
dark matter -- galaxies:haloes 
\end{keywords}



\section{Introduction}

Recent observational studies have provided a new generation of constraints on the amount of small scale structure in the Universe. Two studies of flux anomalies in multiply imaged lensed quasars have inferred the presence of a large number of dark matter subhaloes \citep{gilman19,hsueh19}, and a recent study of Milky Way stellar streams has claimed a similar detection \citep{Banik19}. These studies report the existence of a minimum number of subhaloes in a given mass range around a target host halo -- massive elliptical galaxy haloes for flux anomalies and our Milky way halo in the case of stellar streams -- and can therefore place limits on models of dark matter in which the abundance of haloes is suppressed by the presence of a cut-off in the linear matter power spectrum.

This cut-off can occur in models of sterile neutrino (Ns) dark matter \citep{Dodelson94,Shi99,Laine08,Boyarsky09a,Lovell16} and models with dark radiation interactions in the early Universe \citep{Buckley14,boehm14,CyrRacine16,Schewtschenko16,Vogelsberger16}. These models show a rich phenomenology of matter power spectra, including sharp cut-offs, shallow cut-offs and dark acoustic oscillations (DAOs). This wide variety of power spectrum options is difficult to constrain systematically with any observational probe, including flux anomalies measurements or gaps in stellar streams, and so these observational studies typically instead place constraints on the simple warm dark matter (WDM) thermal relic model first proposed by \citet{Bode01} and later expanded by \citet{Viel05} (hereafter \citetalias{Viel05}). This model contains a single parameter, the thermal relic WDM particle mass $m_\rmn{WDM}$, which is related directly to the half mode wavenumber, $k_\rmn{hm}$, defined as the wavenumber at which the square root of the ratio of the WDM linear power spectrum to the cold dark matter (CDM) linear matter power spectrum -- otherwise known as the transfer function -- is suppressed by a factor of 2. $k_\rmn{hm}$ can be used to define a characteristic mass scale, the half-mode mass, $M_\rmn{hm}$. It is then simple to parametrize the halo mass function through the combination of $M_\rmn{hm}$, a fitting formula and the CDM mass function. The fitting formula used is typically either that derived for field galaxies in trial WDM cosmologies by \citet{Schneider2012} or for the local halo and MW subhalo populations by \citet{Lovell14} (hereafter \citetalias{Lovell14}). Both of which these fits were made to $N$-body simulations that assumed the \citetalias{Viel05} model, and do not reflect the different environments of interest to observational studies, such as low mass field dwarfs and satellites of lensing elliptical galaxies. 

In this {\it Letter} we examine under what conditions two of the approximations outlined above -- the \citetalias{Viel05} approximation to the linear matter power spectrum and the \citetalias{Lovell14} halo fit to the subhalo mass function -- are appropriate fits to the predictions of a set of well-motivated Ns models, and also to the ETHOS model of interacting dark matter that features DAOs \citep{Vogelsberger16}. These predictions are restricted to simple calculations for field haloes; therefore, we will not make comparisons to observational data as these are almost invariably influenced by the abundance and radial distributions of subhaloes. We present our methods and results in Section~\ref{sec:res} and draw conclusions in Section~\ref{sec:conc}.

\section{Method and Results}
\label{sec:res}

We consider four Ns models in this study. Each uses an Ns with a mass of 7.0~keV, and a unique value of the generation mechanism lepton asymmetry, $L_6=8$, 9, 10 and 12; for a discussion of the relationship between $L_6$ and the matter power spectrum properties see \citet{Lovell16}. Throughout this paper we refer to these four models as LA8, LA9, LA10 and LA12 respectively. LA9 is the model with the highest wavenumber cut-off that is consistent with a dark matter decay origin for the 3.55~keV line reported in M31 \citep{Boyarsky14a}, stacks of galaxy clusters \citep{Bulbul14} and the Galactic centre/Milky Way halo \citep{Cappelluti18,Hofmann19}. LA12 is somewhat warmer than the lowest cut-off in agreement with the line and LA10 is an intermediate case. LA8 is the coldest model of any 7~keV Ns, and was used to perform some of the hydrodynamical simulations in \citet{Despali19}, thus we can compare our results to theirs as a check for our method. In addition to these four Ns models we also use the ETHOS4 model of SIDM (hereafter simply `ETHOS'), which was tuned to obtain a rough match between MW satellite simulation predictions and observations \citep{Vogelsberger16}; it was shown in \citet{Lovell18a} that this power spectrum had the same peak wavenumber, $k_\rmn{peak}$, as the 7~keV Ns with $L_{6}\sim9$. 

The momentum distribution functions of the Ns models were computed initially for \cite{Lovell16}, using the computational algorithm of \citet{Laine08}\footnote{An alternative, public algorithm for calculating these momentum distributions was presented by \citet{Venumadhav16}. It was shown briefly in \citet{Despali19} that the \citet{Laine08} computations return higher $k$ cut-offs than the \citet{Venumadhav16} results, and therefore our findings are conservative.}. All five models, plus the CDM counterpart, had their linear matter power spectra computed using the {\sc camb} Boltzmann code, and used cosmological parameters consistent with the \citet{PlanckCP13} results.

\begin{figure}
    \centering
    \includegraphics[scale=0.335]{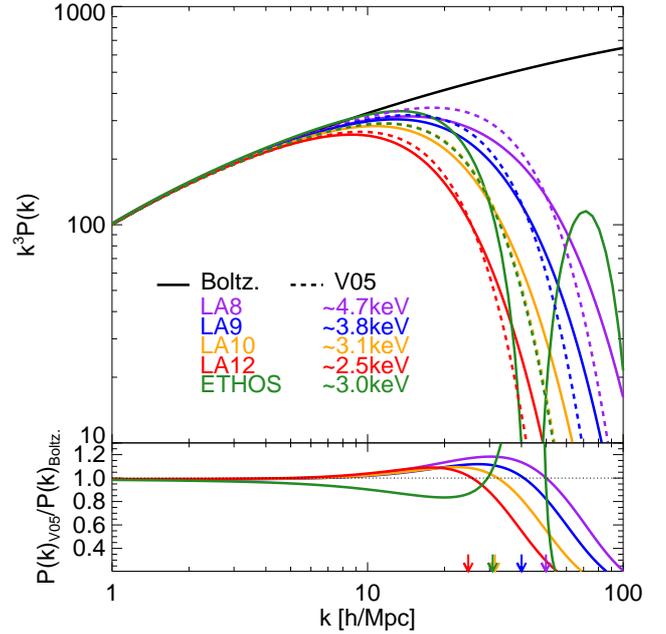}
    \caption{Top panel: the dimensionless matter power spectra for CDM, for four Ns models calculated using the {\sc camb} Boltzmann code, and for their \citetalias{Viel05} equivalents as defined by $k_\rmn{hm}$. The four Ns models are shown as a series of four solid lines: red for LA12, orange for LA10, blue for LA9, and purple for LA8. The ETHOS model is shown as a solid green line. Their \citetalias{Viel05} equivalents, as defined by the $k_\rmn{hm}$, are shown as dashed lines, and CDM as a solid black line. Bottom panel: the ratio between the \citetalias{Viel05} and Boltzmann calculations for each model. The arrows on the $x$-axis mark the values of $k_\rmn{hm}$.  The thermal relic particle masses that correspond to each \citetalias{Viel05} model are given in the figure legend.}
    \label{fig:Pk}
\end{figure}

We discuss an approximation to these linear matter spectra using the \citetalias{Viel05} thermal relic fit, $P_\rmn{WDM}$, which takes the form:

\begin{equation}
    P_\rmn{WDM}(k) = (1+(\alpha k)^{2\nu})^{-10/\nu} P_\rmn{CDM}(k),
\end{equation}

\noindent where $\nu=1.12$ and $P_\rmn{CDM}$ is the CDM power spectrum. $\alpha$ is related to the WDM particle mass as shown by equation~7 of \citetalias{Viel05}, and in turn sets the half-mode wavenumber, $k_\rmn{hm}$, and half-mode mass, $M_\rmn{hm}$, as discussed above.  It is then tempting to use this fit to represent non-thermal relic models for which the fit was not designed. In this study we will test the hypothesis that a \citetalias{Viel05} thermal relic of a given $M_\rmn{hm}$ returns a good approximation to the more complex physics models of Ns and DAOs (hereafter referred to as `Boltzmann-derived'). We note that \citet{Murgia17} have presented a more comprehensive set of fits to these transfer functions, and specifically considered resonantly produced Ns as part of their fitting procedure, but for the purpose of this study we consider only the \citetalias{Viel05} fit as this is the fit most commonly applied in the literature. We therefore calculate $k_\rmn{hm}$ for each of our {\sc camb}-derived power spectra and compute the \citetalias{Viel05} power spectra specified by the same $k_\rmn{hm}$.  As stated above, each \citetalias{Viel05} curve can also be labelled with a thermal relic particle mass: for the LA8, LA9, LA10, LA12 and ETHOS models the \citetalias{Viel05} thermal relic masses are 4.7, 3.8, 3.1, 2.5 and 3.0~keV respectively. We present our results in Fig.~\ref{fig:Pk}. 

Our four Ns Boltzmann-derived power spectra have shallower slopes than their \citetalias{Viel05} counterparts, and this difference correlates with $L_{6}$. The \citetalias{Viel05} fit overestimates the power of the LA12 model for $k<k_\rmn{hm}$ by up to 10~per~cent and the LA8 model by 20~per~cent. It then follows that at scales smaller than $k_\rmn{hm}$ \citetalias{Viel05} progressively underestimates the power, and by a factor of more than two at $2\times k_\rmn{hm}$. ETHOS instead presents a power spectrum cut-off even more abrupt than its \citetalias{Viel05} counterpart, possessing 20~per~cent more power at $k\sim20$~$h/\rmn{Mpc}$ before dropping rapidly. We note that the $k_\rmn{hm}$ of ETHOS is almost identical to that of the LA10 model, even though it has the same $k_\rmn{peak}$ as LA9: we will therefore be able to make a statement about the degree to which $k_\rmn{peak}$ and $k_\rmn{hm}$ influence the mass function.        

The matter power spectra presented here are typically evolved forward in time into halo mass functions using $N$-body simulations of structure formation, from the linear regime to the present day. This is a computationally expensive process for probing a two dimensional parameter space, especially when the target observable is the abundance of dwarf haloes in the local Universe. Only a handful of simulations have been performed of resonantly produced Ns models, almost entirely zoomed simulations of Local Group volumes or individual dwarf haloes \citep{Bozek16,horiuchi2016,Lovell16,Lovell17b,Bozek19,Despali19,Lovell19a} plus the high redshift Lyman-$\alpha$ forest simulations of \citet{Garzilli18}, and are therefore not optimized to compute the average halo mass function.

We therefore adopt three alternative methods: (i) evolving the Boltzmann-derived power spectra forward using the extended Press-Schechter (EPS) method \citep{Press74,Bond91,Benson13}, (ii) repeat the EPS process with the \citetalias{Viel05} counterparts, and (iii) apply the $M_\rmn{hm}$ for each model to the fitting function presented in \citetalias{Lovell14}. For CDM and the Ns models we apply the sharp-$k$ space window filter to obtain the EPS mass functions, whereas for ETHOS we instead apply the smooth $k$-space cut-off introduced by \citet{Sameie19}; we subsequently renormalise the ETHOS mass function to have the same value as CDM at $8\times10^{10}\msun$, in order to compensate for this change in window function. For all five \citetalias{Viel05} fits we use the sharp-$k$ space cut-off. 

We begin our comparison between fits and the Boltzmann code-EPS (B-EPS) results with the \citetalias{Lovell14} halo fits. This fit is given by the \citetalias{Lovell14} equation~7, which we reproduce here:

\begin{equation}
    n_\rmn{WDM}/n_\rmn{CDM}=(1+ M_\rmn{hm}M_\rmn{sub}^{-1})^{\beta},
    \label{eqn:L14}
\end{equation}

\noindent where $n_\rmn{WDM}$ and $n_\rmn{CDM}$ are the WDM and CDM differential mass functions respectively, $M_\rmn{sub}$ is the bound mass of the halo/subhalo as defined by the halo finder and $\beta=-1.3$.  In addition to the \citetalias{Lovell14} fit we compute the mass function fitted by \citet{Despali19} to the subhalo mass functions of a series of four LA8 hydrodynamical simulations, which instead uses the functional form that \citetalias{Lovell14} derived using an extra parameter (their equation~8):

\begin{equation}
    n_\rmn{WDM}/n_\rmn{CDM}=(1+ \gamma M_\rmn{hm}M_\rmn{sub}^{-1})^{\beta},
    \label{eqn:L14-2}
\end{equation}

\noindent originally with $\beta=-1.3$ and $\gamma=2.7$. \citet{Despali19} refit $\gamma=0.35$ for LA8;  this model has $M_\rmn{hm}=1.28\times10^{8}\msun$. 

 The EPS method is typically applied in CDM for the top-hat measure of halo mass, $M_\rmn{TH}$, using the real space top-hat filter. Alternative filters, such as the sharp and smooth $k$-space filters used in this study, are calibrated to approximate roughly the results of the top-hat filter at scales where the different dark matter models are expected to be indistinguishable. We therefore label the masses computed using these two filters with $M_\rmn{TH}$, while bearing in mind that comprehensive comparisons to observations will require a more careful definition of halo mass.

\begin{figure}
    \centering
    \includegraphics[scale=0.33]{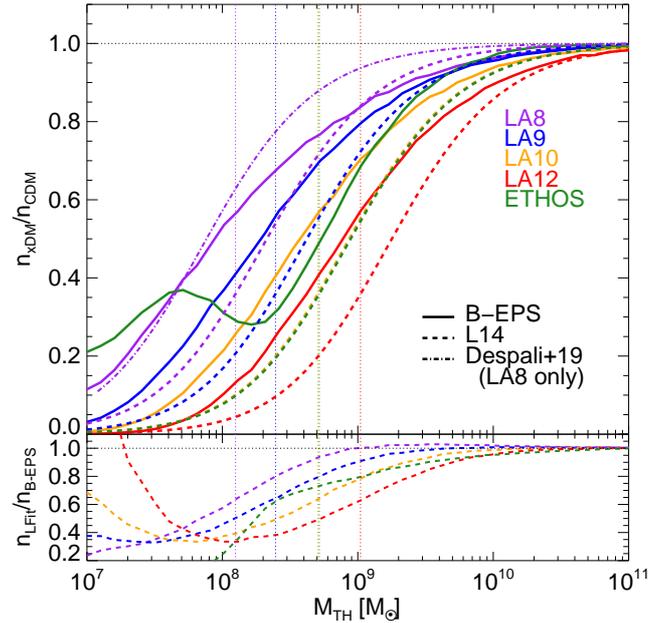}
    \caption{Top panel: the ratio of alternative dark matter halo differential mass functions with respect to CDM. The Ns models are LA8 (purple), LA9 (blue), LA10 (orange) and LA12 (red), and ETHOS is shown in green. The EPS-derived curves are shown as solid lines, and the fit from \citetalias{Lovell14} assuming each model's $k_\rmn{hm}$ as dashed lines. For LA8 we also show the fit to this model produced by \citet{Despali19} as a dot-dashed line. Bottom panel: the ratio of the \citetalias{Lovell14} fits relative to the EPS-derived counterparts. In both panels the vertical dotted lines indicated $M_\rmn{hm}$.}
    \label{fig:MF1}
\end{figure}

One complication to this method is that the \citetalias{Lovell14} fit instead derived for $M_\rmn{sub}$ rather than $M_\rmn{TH}$ or the more common $M_{200}$, defined as the mass contained within the radius of mean density $200\times$ the critical density for collapse, since the latter two measures of mass are not defined for subhaloes. We therefore make the first order assumption that $M_\rmn{200}\approx M_\rmn{sub}$, and then $M_\rmn{TH}=1.2M_{200}$ as is typically found in $N$-body simulations.  We present our results in Fig.~\ref{fig:MF1}, first as ratios with respect to CDM and second as the ratio of the \citetalias{Lovell14} fit to B-EPS.  

There is significant disagreement between the B-EPS and the \citetalias{Lovell14} predictions. The latter predicts 40~per~cent (30~per~cent) fewer haloes at $M_\rmn{hm}$ than do the B-EPS Ns (ETHOS) calculations. This disagreement worsens towards lower masses, particularly for ETHOS as \citetalias{Lovell14} cannot account for the first DAO bump, although at masses below a tenth of $M_\rmn{hm}$ the B-EPS mass function for Ns becomes shallower than the \citetalias{Lovell14} prediction. The \citet{Despali19} fit instead suggests excellent agreement with the LA8 B-EPS results in the $[10^{7},10^{8}]\msun$ mass range crucial for lensing and stream gap studies, at the expense of predicting up to 15~per~cent more haloes than B-EPS at higher masses. Given that the LA8 model is colder than any of the sterile neutrino models that can explain the 3.55~keV line feature, we therefore show that this fit by \citet{Despali19} is the most useful first order fitting function currently available for comparing Ns model predictions with observations, although bespoke simulations will be required in order to make definitive statements about the viability of any particular model. 

\begin{figure}
    \centering
    \includegraphics[scale=0.33]{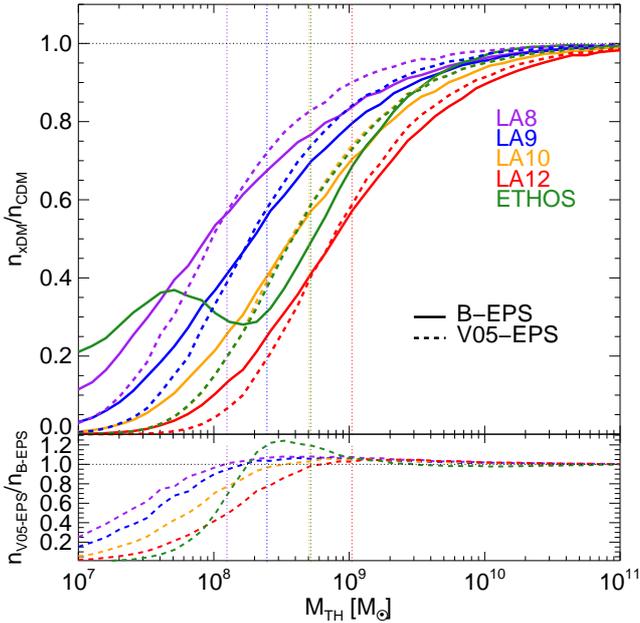}
    \caption{Top panel: as for Fig.~\ref{fig:MF1}, but replacing \citetalias{Lovell14} fits with EPS-derived \citetalias{Viel05} results, shown as dashed lines. Bottom panel: the ratio of V05-EPS to Boltzmann-EPS curves.}
    \label{fig:MF2}
\end{figure}

We repeat this exercise with the \citetalias{Viel05} fits evolved using EPS (V05-EPS) and present the results in Fig.~\ref{fig:MF2}. The agreement between B-EPS and V05-EPS at $M>M_\rmn{hm}$ is better than 10~per~cent, and what discrepancy there is corresponds to an overprediction of haloes as one would expect from the excess power at $k<k_\rmn{hm}$ relative to the Boltzmann calculations (Fig.~\ref{fig:Pk}). The agreement at $M_\rmn{hm}$ itself is much better than 1~per~cent. At lower masses the loss of power in the \citetalias{Viel05} fit is apparent in the over-suppression of haloes, by at least 80~per~cent at $10^{7}\msun$ for all models. The ETHOS V05-EPS fit instead produces 20~per~cent more haloes at $M_\rmn{hm}$ than the B-EPS counterpart, although it inevitably misses the bump due the first DAO at $3\times10^{7}\msun$. We note that the mass function of ETHOS overall bears a stronger affinity to that of the LA10 Ns than that of the LA9, thus we conclude that $k_\rmn{hm}$ is a better predictor of the output mass function than $k_\rmn{peak}$, in so far as one wavenumber is able to specify the entire mass function.

 Finally, as an experiment we have also briefly calculated V05-EPS fits that are specified at two alternative positions on the matter power spectrum, where $\sqrt{P_\rmn{WDM}/P_\rmn{CDM}}=0.25$ and 0.75, compared to $\sqrt{P_\rmn{WDM}/P_\rmn{CDM}}=0.5$ for the half mode mass; we label these two wavenumbers $k_{0.25}$ and $k_{0.75}$ respectively. Using $k_{0.75}$ to specify the fit shifts the point of divergence from the B-EPS result to higher masses than is the case for $k_\rmn{hm}$, and therefore offers a worse fit at $M_\rmn{TH}<10^{8}\msun$ (not shown). Taking $k_{0.25}$ instead overpredicts the number of B-EPS haloes by 20~per~cent at $M_\rmn{hm}$ while still underpredicting the number of $10^7\msun$ haloes by at least 40~per~cent. We therefore conclude that $k_\rmn{hm}$ remains the most appropriate option for choosing a \citetalias{Viel05} fit, and future studies should focus on more detailed fitting forms such as presented by \citet{Murgia17}.       

\section{Conclusions}
\label{sec:conc}

Recent observational studies of lensing flux anomalies \citep{gilman19,hsueh19} and stellar stream gaps \citep{Banik19} have reported strong constraints on the properties of dark matter, including the presence or otherwise of a matter power spectrum cut-off predicted by well-motivated particle physics models. These studies are constrained to test single parameter models and it has therefore not been clear whether the underlying particle physics models, in their full complexity, are in tension with the data. 

In this {\it Letter} we have used some simple analyses of sterile neutrino (Ns) and dark acoustic oscillations (DAO, specifically ETHOS) to test some of the fits found in the literature. We have found that the \citet{Viel05} (\citetalias{Viel05}) counterpart to the linear matter spectrum, as defined by the half-mode wavenumber $k_\rmn{hm}$ is accurate to within 10~per~cent for wavenumbers $k<30$~$h/$Mpc for Ns models consistent with being the origin of the reported 3.55~keV line \citep{Boyarsky14a,Bulbul14}, although the coldest model (LA8) is overpredicted by up to 20~per~cent (Fig.~\ref{fig:Pk}). At much smaller scales the fit progressively underpredicts power spectrum. The ETHOS model is instead  steeper than its \citetalias{Viel05} counterpart.    

We then computed $z=0$ halo mass functions using three methods: the fitting function of \citet{Lovell14} (\citetalias{Lovell14}) assuming the half-mode mass $M_\rmn{hm}$ associated with each matter power spectrum, applying the extended Press-Schechter (EPS) formalism to the \citetalias{Viel05} fits (V05-EPS), and again applying the EPS formalism to the original Boltzmann code-derived power spectra (B-EPS). We showed that the \citetalias{Lovell14} fit applied to this scenario underpredicts the number of haloes at $M_\rmn{hm}$ relative to the B-EPS measurement, whereas V05-EPS agrees with B-EPS to better than 5~per~cent at the same mass. However, in the $[10^{7},10^{8}]\msun$ band that is crucial for lensing and stream gap studies, both \citetalias{Lovell14} and V05-EPS  underestimate the amount of substructure in all five models, relative to the B-EPS calculation, by at least 50~per~cent over much of that range. It is therefore crucial to determine accurate models, whether improved fits that are designed to accommodate resonantly produced Ns models \citep{Boyarsky2009c,Murgia17} or to use EPS models / $N$-body models, when determining whether a given model has been ruled out. 

We caution that attempts to constrain these models must account for other phenomena that we do not consider here. These include the different spatial distribution of WDM subhaloes within a host halo compared to CDM \citep{Lovell14,Bose17a,Despali19} and likely altered destruction rates due to the lower concentrations of WDM haloes. One must account for the different definitions of the halo / subhalo mass, especially when subhaloes within a host halo are expected to contribute a significant part of the signal, and model the impact of baryon physics on any change to the mass. It will therefore be imperative for studies that place competitive constraints on generic matter power spectrum parameters to be followed up with dedicated simulations, that will in turn ascertain whether a given model has been ruled out.

\section*{Acknowledgments}

MRL would like to thank Jes\'us Zavala for useful conversations and comments on the text, and to thank the anonymous referee for useful comments. MRL is supported by a Grant of Excellence from the Icelandic Research Fund (grant number 173929).

\bibliographystyle{mnras}

\bsp	
\label{lastpage}
\end{document}